\documentstyle[pre,aps,epsfig,array]{revtex}
\begin{document}
\draft

\twocolumn[\hsize\textwidth\columnwidth\hsize\csname@twocolumnfalse\endcsname

\title{Mean-field approximation for a limit order driven market model
}

\author{Franti\v{s}ek Slanina}
\address{        Institute of Physics,
	Academy of Sciences of the Czech Republic,\\
	Na~Slovance~2, CZ-18221~Praha,
	Czech Republic\\
        e-mail: slanina@fzu.cz
}
\maketitle
\begin{abstract}
The mean-field variant of the model of limit order driven market
introduced recently by Maslov is formulated and solved. 
The agents do not have any strategies and the memory of the system
is kept within the order book.
We show that he evolution of the order book is governed by a matrix
multiplicative process. The resulting stationary distribution of
step-to-step price changes is calculated. It exhibits a power-law tail
with exponent 2. We obtain also the price autocorrelation function,
which agrees qualitatively with the 
experimentally observed negative autocorrelation
for short times. 
 \end{abstract}

\twocolumn]

\section{Introduction}

The complexity of market behavior, seen as a particular example of
a natural phenomenon, fascinated physicists since long time ago
\cite{an_ar_pi_88}. 
The main source of interest comes from a kind of critical behavior,
made explicit by the power-law distribution and scaling in the
economic time series, first observed by Mandelbrot
(see \cite{mandelbrot_99} and references therein) and studied in 
detail by Mantegna and  Stanley \cite{mantegna_91,ma_sta_95,ma_sta_99}
and subsequently by many 
others (see e. g. \cite{bou_pot_00,ga_ca_ma_zha_97,va_au_98,go_me_am_sta_98,ple_gop_am_mey_sta_99,go_ple_am_me_sta_99}). 

The scaling and multifractal properties call for an explanation in
terms of a model mimicking the behavior of individual agents in the
market. The physicist's optimism in looking for such a model might be
strengthened by recent successes in modeling other social phenomena,
e. g. in the cellular-automata models
of traffic \cite{na_sch_92}.  
The idea consists in assuming that all the overwhelming complexity of
a human being is irrelevant in certain special conditions: when
driving a car, only a very basic set of behaviors is at work. 
Similarly, it is assumed that a trading agent, when put on the floor,
follows only a limited set of instincts or acquired
patterns. Therefore, in this approach the economic complexity is not
due to the 
intrinsic complexity of each agent (as usual hand-waving arguments by
liberal opponents of ``reductionism'' state), but an emergent property
of a large set of non-linearly interacting simple units. 

Many microscopic stock market models emerged during last several
years. One of the first ones was the model
introduced by
Levy, Levy and Solomon 
\cite{lev_lev_sol_94,le_le_so_95,lev_sol_96a}
which captures essential features of the price fluctuations and
explains also the power law distribution of investor's wealth, which
is the famous Pareto law.

Another approach was used in the model of 
Bak, Paczuski and Shubik \cite{ba_pa_shu_97}: buyers and sellers are 
represented by particles subject to a reaction-diffusion
process. Introduction of a non-trivial strategy of the agents lead to
a realistic value of the Hurst exponent for the price fluctuations.
The simpler version of the model was then solved analytically
\cite{ta_tia_99}.
Variety of other approaches were investigated \cite{ta_mi_hi_ha_92,sa_ta_98,lux_mar_99,co_bou_97,ca_ma_zha_97,cho_sta_99,so_sta_tak_99,bou_co_98,iori_99,iori_00a,sla_zha_99,sla_zha_01,kul_ker_01,cha_zha_97,cha_ma_zha_99,jo_ha_hu_zhe_99,cha_mar_zha_01a}.

The above mentioned models mostly do not take into account the 
realistic details of the price formation through the book of
orders. This mechanism was implemented in the model set up by
Maslov \cite{maslov_99} and a similar perspective was then used in a recent
series of papers by Matassini and Franci
\cite{fra_mat_00,mat_fra_01,mat_fra_01a}. 
The book of orders perspective to market modeling was empirically investigated 
in \cite{mas_mil_01,cha_sti_01}.

With such a diversity of models, most of which giving plausible
explanation of observed facts a question arises, whether there is a
common mechanism behind various approaches, making them essentially
equivalent. Indeed, it was found that such a mechanism may be the
multiplicative stochastic process repelled from zero, or
multiplicative-additive process. It was studied thoroughly by various
authors and in diverse contexts
\cite{lev_sol_96a,deutsch_94,lev_sol_96,ta_sa_ta_97,sornette_97a,so_co_97,sornette_98c,bi_ma_le_so_98,ma_mas_zha_98,solomon_99,hua_sol_00,sol_lev_00,bou_mez_00}.
The goal we pose in this work is to show that essentially the same
mechanism  is responsible for the power-law distribution of price
changes also in the limit-order model.

\section{Limit order driven market model}
Recently Maslov \cite{maslov_99,mas_mil_01} proposed a model, 
based on the
assumption that
there are two kinds of market participants. Prudent investors place
their orders at prescribed price and a trade occurs as soon as there
is anybody accepting that price. On the other hand, speculators buy
and sell at any moment at the price which is available in the
market. The price signal $p(t)$ was found to have power-law spectrum, with
Hurst exponent $H=1/4$. The price changes during unit time interval
$x=p(t+1)-p(t)$ have 
probability distribution which follow clear power law $P(x)\sim
x^{-(1+\alpha)}$ in two regimes. For small $x$  
the  exponent is $1+\alpha_1=0.6\pm 0.1$, while in the regime of large price
changes the exponent is $1+\alpha_2=3\pm 0.2$. These values are to be
compared to the experimentally found $1+\alpha_1\simeq 2.5$
 and $1+\alpha_2\simeq 4$ respectively \cite{bou_pot_00,go_me_am_sta_98,ple_gop_am_mey_sta_99,go_ple_am_me_sta_99}.  

The model \cite{maslov_99} can be described as follows. There are
orders to buy and sell placed on a straight
line, which is the axis of the price $x$. In stable situation, all buy
orders are lower than all sell orders, so that we can describe the
state by single function $\rho(x)$, density of the orders, and a
number $\xi$ which is the last realized price. Then, all $x<\xi$
correspond to buy, all $x>\xi$ to sell orders. 

Two events can change the state. First, new limit orders may be dropped,
such that $\rho(x)\to\rho(x)+\eta(x-\xi)$ We suppose the function
$\eta(x)$ equal in all events and symmetric, $\eta(x)=\eta(-x)$. 
Second, a market-price order can arrive. An order to buy an amount $s$
results to clearing all sell orders up to the price $\xi+x_+$, where 
\begin{equation}
\int_\xi^{\xi+x_+}\rho(x){\rm d}x = s
\end{equation}
The new price is then $\xi\to\xi+x_+$, so that $x_+$ is the price
increment, while the new density is 
$\rho(x)\to(1-\theta(x-\xi)\theta(\xi+x_+-x))\;\rho(x)$. Analogical
formulae will hold for the sell order.

As we can see, there are no strategies which would lead the agents to
perform specific actions. The model is barely stochastic. The
long-term memory of the system and thus a possible power-law behavior
stems from the order book, or the time-dependent density function
$\rho(x)$ which may keep arbitrarily old orders.

\section{Matrix formulation}

Our essential approximation to this model will consist in supposing uniform 
density of orders on each side from the current price level $\xi$. 
In reality, both dropping new limit orders and clearing them by market
orders makes the density of states uneven and fluctuating. When
supposing that after an event the uniform density of states is
restored, we make a kind of ``mean-field'' approximation: the actual
position of each limit order is not important, as if they were freely
moving particles making an effective medium, within which the
price fluctuates. High density of the medium will result in smaller
price fluctuations and vice versa.

The
density on the upper side will be denoted $\rho_+$, on the lower side
$\rho_-$. It is convenient to describe the densities in terms of the
potential price changes, which would occur if a market-price order
arrives. They are simply $x_+=s/\rho_+$ for buy and $x_-=s/\rho_-$ for
sell order. The
numbers $x_\pm$ form a vector 
$X=\left(\begin{array}{l}
x_+\\
x_-
\end{array}
\right)$ which
performs a stochastic process, as the densities $\rho_\pm$ and
therefore the numbers $x_\pm$ are updated after arrival of each
order. 
The dynamical rules of the process represent a
simplified version of the limit order driven dynamics. 

There are be
three types of events: (i) dropping of limit orders (ii) market-price
order to buy, and (iii) market-price order to sell. We suppose that all
market price orders have the same volume $s$ and all limit-order
events the same volume $v$. Further we assume that market orders to
buy and sell occur with the same probability. In order to keep the
total number of limit orders constant on average, we should suppose
that at given moment there is probability $p=s/(s+v)$ to drop a
limit orders. Each of the market-price events (ii) and (iii)  have
then equal probability $(1-p)/2$. (Here we tacitly assume that the
limit orders which are not met last forever. One can also investigate
more realistic variants, where the limit orders slowly die out).
 
Let us 
investigate first the consequence of an arrival of a market-price
order to buy. By definition, the price level increases by
$x_+$. As the density 
of orders is constant, average density on the right-hand side from the
new price is unchanged. If now another buy order arrives, it finds the
same density and the price change is the same too. 
Therefore, the new value of $x_+$ is equal to the old one,
$x_+\to x_+$. On the other hand, if now a new market order to sell
arrives, there are no limit orders in the interval of width $x_+$
below the current price level, and when we go further down, there is a
constant density $s/x_-$. As a result, the price decreases by
$x_++x_-$. Hence, the new value of $x_-$ is $x_-\to
x_++x_-$. 

To sum it up,
the effect of the buy order consists in the replacement 
\begin{equation}
\begin{array}{l}
x_+\to x_+\\
x_-\to
x_++x_-\quad .
\end{array}
\end{equation}
It
can be expressed in matrix form
\begin{equation}
X\to X^\prime = T_+ X
\end{equation}
where
\begin{equation}
T_+ =\left( 
\begin{array}{ll}
1&0\\
1&1
\end{array}
\right)\; .
\end{equation}
Similarly, for the action of a sell order we get
\begin{equation}
X\to X^\prime = T_- X
\end{equation}
where
\begin{equation}
T_- =\left( 
\begin{array}{ll}
1&1\\
0&1
\end{array}
\right)\; .
\end{equation}

Now we turn to the changes due to dropping limit orders.
It is necessary to specify the function $\eta(x)$, representing the
average volume of orders set at distance $x$ from the current price.  
As we already
mentioned, we suppose it to be an even function. Moreover, the volume
was supposed 
to be fixed, $\int\eta(x){\rm d}x = v$. We apply here the simplest
choice $\eta(x)=\frac{v}{2}(\delta(x-d)+\delta(x+d))$, which means
that all new orders are placed at the same distance $d$ from the
current price, either below (buy) or above (sell). This
distribution reflects the fact, that the limit orders are not
typically set arbitrarily close to the current price, but there is
a certain minimum offset $d$. 

Dropping limit orders affects the vector $X$ according to the formula
\begin{equation}
x_\pm\to\frac{1}{2}(3-1/p)\,x_\pm\quad .
\label{eq:changexlimit}
\end{equation}
 Indeed, a buy (or sell) order
will annihilate the amount $v/2$ from the just deposited limit order
and amount $s-v/2$ from the original density of old limit
orders. The shift is therefore $x_\pm=(s-v/2)/\rho_\pm$. Writing $v$
in terms of the probability $p$, i. e. $v=(1/p-1)s$, we obtain the
formula (\ref{eq:changexlimit}).
So,
in matrix form 
we have 
\begin{equation}
X\to X^\prime = S X
\end{equation}
where
\begin{equation}
S =\frac{1}{2}(3-\frac{1}{p})\left( 
\begin{array}{ll}
1&0\\
0&1
\end{array}
\right)\; .
\end{equation}

The price changes only after a market order is issued, while dropping
limit orders leaves the price unchanged. So, between two 
subsequent shifts of the price, $m\ge 0$ limit orders can arrive, with
probability $P_m(m)=(1-p)p^m$. The change of the vector $X$ due to one
market order and $m$ limit orders is $X\to S^m\,T_\pm\,X$. When
calculating the evolution of the probability distribution for $X$, we
should sum over all possible realizations.
Hence, the probability distribution for
the vector $X$ should  satisfy the equation
\begin{eqnarray}
P_X&&(X)=\\
&&\frac{1}{2}\sum_{\sigma=\pm}
\sum_{m=0}^\infty
\int{\rm d}X^\prime\,P_m(m)P_X(X^\prime)\delta(X-S^m \, T_\sigma \,X^\prime)
\nonumber
\end{eqnarray}
in the stationary state. 

\section{Distribution of price changes}

We will make a further approximation at this stage. The matrix $S$ is
simply a unit matrix multiplied by a constant. If the same were true
also for the matrices $T_\pm$, the process would be reduced to a simple
multiplicative random walk, whose properties are well known
and their relevance in modeling price fluctuations is testified by a
series of models, as mentioned in the Introduction.

Our approximation will consist first in replacing the matrices $T_\pm$
by the average $\overline{T}=\frac{1}{2}(T_++T_-)$ and furthermore, we
will take only the highest eigenvalue of the matrix $\overline{T}$,
which is $3/2$. Then, instead of a pair of price changes $x_+$ and
$x_-$ we have a single scalar quantity $x$, describing the absolute
value of the price change. 

Note that the same results can be obtained by assuming from beginning,
that $x_+=x_-$, i.e. that the density of states is equal on both sides
of the price level. This means, that we make a further ``mean-field''
approximation, suppressing not only the fluctuations along the price
axis, but also fluctuations from one side to the other of the already
averaged density of states.

This way we define our multiplicative random process. The fact that
there is a small but finite offset $d$ in placing the limit orders
ensures that the values of $x_\pm$ (therefore also of $x$) cannot be
smaller than $d$. This feature plays the role of ``repulsion from
zero'', which was found essential for establishing the power-law tails
\cite{lev_sol_96,so_co_97} and is usually guaranteed by the additive
term \cite{ta_sa_ta_97,sornette_98c}.

For the probability distribution of the price changes we obtain
\begin{eqnarray}
P(x)&&=\sum_{m=0}^\infty(1-p)p^m\times\label{eq:forP}\\
&&\times\int{\rm d}x^\prime\,P(x^\prime)\,
\delta\left(x-\frac{3}{2}\,x^\prime\,\left(\frac{3-\frac{1}{p}}{2}\right)^m
\right)  
\nonumber
\end{eqnarray}
and assuming a power-law tail of the probability distribution in the
form $P(x)\sim x^{-1-\alpha}$
we obtain the following equation for the exponent
\begin{equation}
\sum_{m=0}^\infty(1-p)p^m
\left(\frac{3}{2}\left(\frac{3-\frac{1}{p}}{2}\right)^m\right)^\alpha=1\;.
\end{equation}

As can be easily checked, apart from the trivial solution $\alpha=0$
it has a non-trivial solution $\alpha=1$, independent of $p$.
Therefore, the distribution of price changes has a power-law tail
\begin{equation}
P(x)\simeq x^{-2}\;\; .
\end{equation}

Note that the calculation could be further simplified by writing the
equation analogical to (\ref{eq:forP}), relating the probability
distribution just after single step. Then,
 instead of the sequence of steps consisting of 
one market order followed by $m$ limit orders
we have one step being either limit
or market order. This equation gives precisely the same power-law
tail. However, such an approach is slightly inconsistent, because
setting a limit order does not imply any trade, 
thus the price change at this moment is zero.

\section{Price autocorrelation function}

One of the well-known facts about financial data series is the
negative short-time autocorrelation of price changes
\cite{bou_pot_00}. Here we will show, how this effect naturally
emerges from the matrix nature of our stochastic process.

We will compute the autocorrelation function defined as
\begin{equation}
C(t,t+\tau)=\frac{\langle x(t)x(t+\tau)\rangle}{\sqrt{\langle x^2(t)\rangle \langle x^2(t+\tau)\rangle }}
\end{equation}
where $x(t)$ is the actual price change at time $t$ and $\tau\ge
1$. We will denote 
$M(t)\in\{ S,T_+,T_-\}$ the matrix describing the action
performed at time $t$ and $P_M(M)$ its probability
distribution. Of course, we introduced already $P_M(T_-)= 
P_M(T_+)=(1-p)/2$ and $P_M(S)=p$. 

Now we introduce the function of taking the price change from the
vector $X={x_+ \choose x_-}$
\begin{eqnarray}
{\cal X}(X;M)=&x_+ &\;\;{\rm if}\;M=T_+\nonumber\\
=&0 &\;\;{\rm if}\;M=S\label{eq:defC}\\
=&x_- &\;\;{\rm if}\;M=T_-\nonumber\; .
\end{eqnarray}
Note that the operator ${\cal X}$ is linear in the argument $X$.
Then
\begin{equation}
\langle x(t)x(t+\tau)\rangle
=\int{\rm d}X\,P_X(X)\widetilde{C}(X)
\end{equation}
where
\begin{eqnarray}
&&\widetilde{C}(X)=\sum_{M(t)}...\sum_{M(t+\tau)}\prod_{i=0}^\tau\,
P_M(M(t+i))\times\\
&&\times\,{\cal X}(X;M(t))\,
{\cal X}(M(t+\tau-1)...M(t)X;M(t+\tau))\nonumber
\end{eqnarray}

We find easily, using the linearity if the operator ${\cal X}$ and
introducing the sign vector $E=(1,-1)$
\begin{eqnarray}
\widetilde{C}(X)=\frac{1-p}{2}E&&\left(\sum_M
P_M(M)M\right)^{\tau-1}\times\\
&&\times(x_+T_+ -
x_-T_-)X\;\; .
\nonumber
\end{eqnarray}
The multiplication by $E$ extracts only the lower one of the two
eigenvalues of the averaged matrix
\begin{equation}
\sum_M P_M(M)M =\frac{1}{2}
\left(\begin{array}{ll}
1+p&1-p\\
1-p&1+p
\end{array}
\right)\;\; .
\end{equation}
The lower eigenvalue is $p$, hence
\begin{equation}
\widetilde{C}(X)= - (1-p)p^{\tau-1}x_+x_-\;\; .
\end{equation}

We can calculate similarly the corresponding expression for the
denominator of the equation (\ref{eq:defC}). We obtain at the end
\begin{equation}
C(t,t+\tau)=-p^{\tau-1}\frac{2\langle x_+x_-\rangle}{\langle
x_+^2\rangle  + \langle x_-^2\rangle}\;\; .
\label{eq:forC}
\end{equation}

We can clearly observe the negative autocorrelation which decays with
characteristic time which depends on the relative frequency of putting
the limit and market orders, measured by the probability $p$.
The result obtained suffer from the divergence of second moments of
the variables $x_+$, $x_-$, resulting from the power-law tail
calculated in the last section.
However, only the ratio of the moments enter the formula
(\ref{eq:forC}). Moreover, if we suppose as an initial condition 
a distribution for $x_+$, $x_-$ with finite moments, the moments will
remain finite for any finite time and we naturally expect that the
ratio of the second moments will converge to a finite value even if the
second moments themselves diverge.

\section{Conclusions}

In conclusion, we solved in the mean-field approximation the
Maslov model of stock market fluctuations. We found a stationary
distribution of price changes with a power law tail with  the exponent
$1+\alpha=2$, which is within the L\'evy stable region. We found
negative short-time autocorrelation of the price changes, decaying
exponentially with time. The relaxation time depends on the relative
frequency of putting market orders and limit orders: the decay is
slower if a market order comes only after more limit orders. This is
intuitively clear, because it is the market order that ensures the
liquidity. 

Our result differs in two important points from the simulations of
Maslov \cite{maslov_99}. First, the numerical value of the power-law
tail exponent is different. This can be attributed to the
approximation we made in the form of the density of orders
$\rho(x)$. Indeed, we assumed, as a zero approximation, constant
density. On the other hand, it is known from the solution of the
reaction-diffusion model of market \cite{ta_tia_99} that the density
may have complicated non-trivial form. Another source of the
difference may be the neglect of fluctuations. 

Another difference consist in lacking the second power-law regime for
small price changes. However, as discussed in \cite{maslov_99}, this
second and different power law comes from the fact, that the new limit
orders may be placed farther than the reach of the price change. As we
implicitly supposed that the new orders are put to very small distance 
$d$ from the current price, we can observe only the distribution for
price changes larger than $d$.

\acknowledgments{I am  indebted to Y.-C. Zhang and the University of Fribourg,
Switzerland, for financial support and kind hospitality.
I wish to thank Sergei Maslov for stimulating discussions and many
claryfying remarks.
This work was supported by the Grant Agency of the Czech Republic,
grant project No. 202/01/1091.}

 
\end{document}